\documentclass[a4paper,11pt]{article}
\usepackage{pos}
\usepackage{float}

\title{Three-dimensional Gross-Neveu model with two flavors of staggered fermions}

\author*[a]{Sandip Maiti}
\author[a]{Debasish Banerjee}
\author[b]{Shailesh Chandrasekharan}
\author[c]{Marina Krstic Marinkovic}
\affiliation[a]{Saha Institute of Nuclear Physics,\\
  HBNI, 1/AF Bidhannagar, Kolkata 700064, India}

\affiliation[b]{Duke, University,\\
Department of Physics, Box 90305, Durham, NC 27708, USA}

\affiliation[c]{ETH Zurich,\\
Institute for Theoretical Physics, ETH Zurich, 8093 Z\"urich, Switzerland}

\emailAdd{sandip.maiti@saha.ac.in}
\emailAdd{debasish.banerjee@saha.ac.in}
\emailAdd{sch27@duke.edu}
\emailAdd{marinama@phys.ethz.ch}

\abstract{We introduce a strongly interacting lattice field theory model containing two flavors
of massless staggered fermions with two kind of interactions: (1) a lattice current-current
interaction, and (2) an on-site four-fermion interaction. At weak couplings, we expect a
massless fermion phase since our interactions become irrelevant at long distances. At strong
couplings, based on previous studies, we argue that our lattice model contains two different 
massive fermion phases with different mechanisms of fermion mass generation. In one phase, 
fermions become massive through Spontaneous Symmetry Breaking (SSB) via the formation of a 
fermion bilinear condensate. In the other phase, fermion mass arises through a more exotic 
mechanism without the formation of any fermion bilinear condensate. Our lattice model is free 
of sign problems and can be studied using the fermion bag algorithm. The longer term goal 
here is to study both these mass generation phenomena in a single model and understand how 
different phases come together.}

\FullConference{%
 The 38th International Symposium on Lattice Field Theory, LATTICE2021
  26th-30th July, 2021
  Zoom/Gather@Massachusetts Institute of Technology
}

\begin{document}
\maketitle
\section{Introduction}
The traditional approach for massless fermions to acquire mass is through the formation of
fermion bilinear condensates. In the standard model of particle physics, this occurs through
the mechanism of spontaneous symmetry breaking (SSB), which gives masses to quarks and
leptons \cite{Logan:2014jla}. In the electro-weak sector of the Standard Model, the Higgs field
plays a central role in generating  masses for fermions through Yukawa couplings \cite{Pich:2012sx}. 
Due to SSB of the electro-weak symmetry, the Higgs field gets an expectation value, which gives 
rise to fermion bilinear mass terms in the low energy effective theory. In the strong interaction
sector, spontaneous breakdown of chiral symmetries is responsible to create a fermion bilinear
quark condensate. On the other hand, studies of lattice Yukawa models have found exotic phases
at strong couplings \cite{AOKI1992229,LEE1990265}, where fermions seem to acquire mass without 
breaking of any symmetries of the theory. Such phases have been called the paramagnetic strong 
phase (PMS-phase) to contrast it with the phase at weak couplings where fermions remain massless due
to lack of any symmetry breaking, which is referred to as the paramagnetic weak phase
(PMW-phase). In the PMS phase, all fermion bilinear condensates vanish, but fermions are still
massive \cite{Catterall:2015zua,Hasenfratz:1989jr}. The traditional spontaneously broken phase 
is called the ferromagnetic phase (FM-phase).

In this work, we introduce a new lattice field theory model with four-fermion interactions to
explore all these three phases under one framework. An important feature of our model is that
it is free from sign problems and can be studied using the fermion bag algorithms without
introducing a bare mass for the fermions \cite{Chandrasekharan:2009wc}. We focus on $2+1$ 
dimensions, since the four-fermion interactions are known to produce many interesting quantum 
critical points. For example, in the so-called lattice Thirring model with a current-current 
interaction $U$, an interesting second order phase transition between the PMW and FM phases 
was observed even with a single flavor of staggered fermions 
\cite{Chandrasekharan:2011mn, Chandrasekharan:2013aya}. With two flavors of staggered fermions
but with a single site four-fermion interaction $U^\prime$, an exotic second order transition
between the PMW and PMS phases was observed \cite{Ayyar:2014eua,Chandrasekharan:2013aya}. Our
model brings together both of these transitions into a single phase diagram by introducing a
model with both interactions $U$ and $U^\prime$, thus giving us a way to study how the two
mechanisms for fermion mass generation can come together and if there are interesting new phase
transitions between them. 

\begin{figure}[!tbh]
	\centering	
	\includegraphics[width=0.9\textwidth]{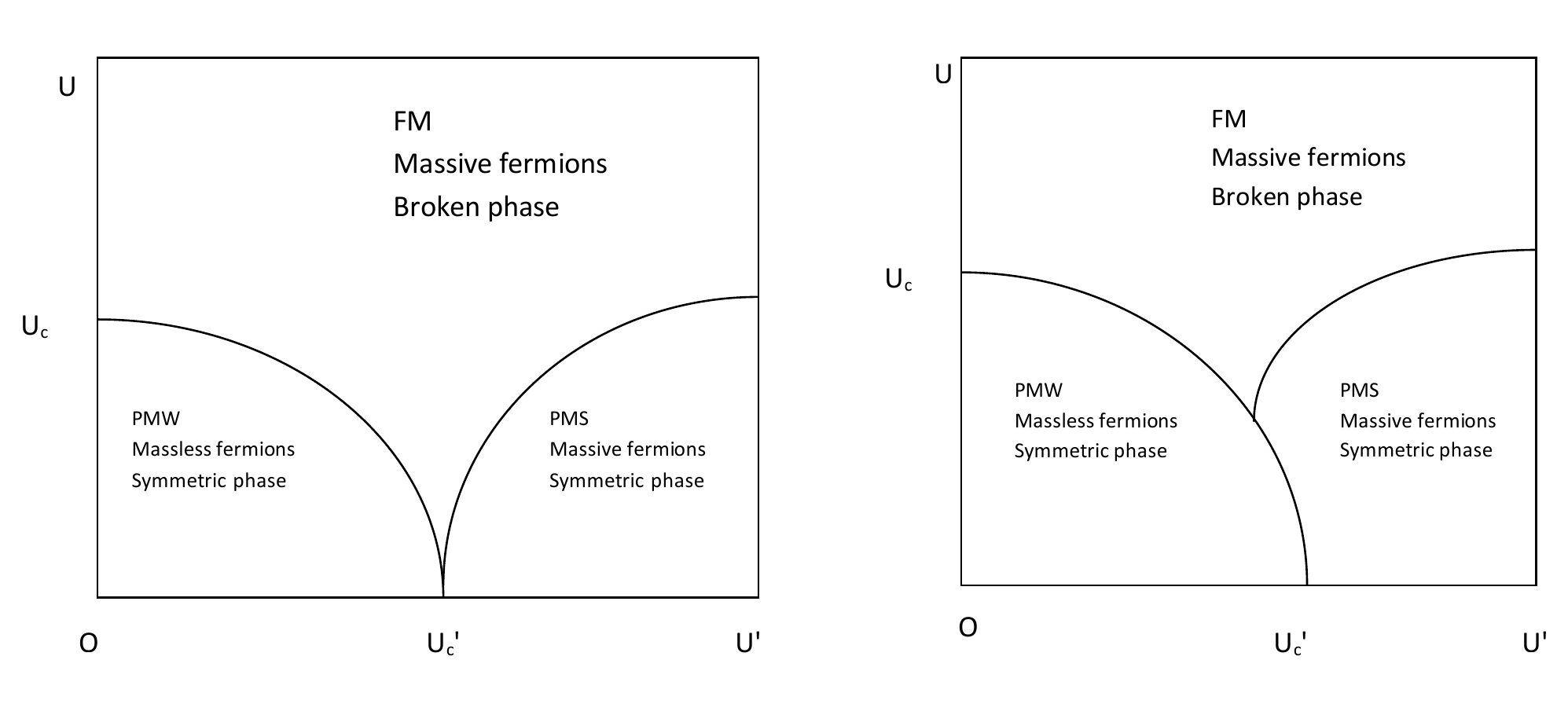}
	\caption{\label{fig:pd}Two possible phase diagrams of our model, containing two types of
	interactions $U$ and $U'$.}
\end{figure}

The possible phase diagrams in our model are shown in Fig.~\ref{fig:pd}. The left phase diagram
implies that $U=0$ is a special axis where, due to enhanced symmetries, a direct second order
phase transition between PMW and PMS phase was possible. Since this generically not possible, 
introduction of $U$ introduces an intermediate phase. On the other hand, the right phase diagram 
implies that the direct phase transition between the PMW and PMS phases is more robust and is 
not a consequence of the special symmetries of the $U=0$ axis. Other interesting questions are 
related to the nature of the phase transitions between the PMW and FM phases and the PMS and FM 
phases. If they are second order, we can compute the critical exponents of these transitions. 
The nature of the multi-critical point where all the three phases meet is another possible 
subject for investigation.

\section{Model}
We study a $2+1$ dimensional lattice model containing two flavors of massless staggered
fermions interacting with each other via four-fermion interactions. The Euclidean action of our
model is given as a sum of three terms $S = S_0 + S^{\text{int}}_1 + S^{\text{int}}_2$, where
$S_0$ denotes the free staggered fermion action for each of the flavors $u$ and $d$ and is
given by
\begin{align}
    S_0 &= \sum_{\langle x,y \rangle} \left(\bar{u}_x M_{x,y} u_y + \bar{d}_x M_{x,y} d_y \right),
\end{align}
$S^{\text{int}}_1$ is a nearest neighbor current-current interaction term within each flavor
\begin{align}
    S^{\text{int}}_1 &= -U \sum_{\langle x,y \rangle} \left( \bar{u}_x u_x \bar{u}_y u_y + 
    \bar{d}_x d_x \bar{d}_y d_y \right),    
\end{align}
and $S^{\text{int}}_2$ is a single site interaction between the two flavors given by
\begin{align}
    S^{\text{int}}_2 &= -U^\prime \sum_{x} \left( \bar{u}_x u_x \bar{d}_x d_x \right).
\end{align}
The matrix $M$ is the massless staggered fermion matrix defined by
\begin{align}
  M_{x,y} &= \sum\limits_{\hat{\mu}} \frac{\eta_{x,\hat{\mu}}}{2}(\delta_{x+\hat{\mu},y} -
  \delta_{x-\hat{\mu},y}),
\end{align}
where $x \equiv (x_0,x_1,x_2)$ denotes a lattice site on a 3 dimensional cubic lattice and
$\hat{\mu}$ represent unit vectors in the three directions. The $\eta_{x,\hat{\mu}}$ are the usual
staggered phase factors defined as: $\eta_{x,\hat{0}}=1$, $\eta_{x,\hat{1}}=(-1)^{x_0}$ and 
$\eta_{x,\hat{2}}=(-1)^{x_0+x_1}$. When $U^\prime=0$, the action $S$ represents two decoupled 
single flavored lattice Thirring models that has been studied earlier \cite{Chandrasekharan:2011mn}. 
The coupling $U^\prime$ couples the two flavors with each other. When $U=0$, the lattice model 
is the same as the one that was studied in Ref.~\cite{Ayyar:2014eua}.

Our lattice model has a rich internal symmetry structure. The free fermion action $S_0$ is symmetric
under $SU(4)\times U(1)$ transformations, while $S^{\text{int}}_1$ is invariant under
$SU(2)\times U(1) \times SU(2)\times U(1)$ and $S^{\text{int}}_2$ is symmetric under $SU(4)$
transformation. Hence, the action $S$ of our model is invariant under $SU(2)\times SU(2) \times
U(1)$ transformations. The observables we wish to measure include local four-fermion condensates given by
\begin{align}
N_u = \frac{2}{V} \sum_{\langle x,y\rangle} \Big\langle \bar{u}_x u_x \bar{u}_y u_y \Big\rangle,\quad
N_d = \frac{2}{V} \sum_{\langle x,y\rangle} \left\langle \bar d_x d_x \bar d_y d_y \right\rangle,\quad
N_i = \frac{1}{V} \sum_x \left\langle \bar u_x u_x \bar d_x d_x \right\rangle,
\end{align}
where the sum $\langle x,y\rangle$ is over nearest neighbor bonds. Similarly, fermion bilinear 
susceptibilities given by
\begin{align}
    \chi_{uu} = \frac{1}{V} \sum_{x,y} \Big\langle \bar{u}_x u_x \bar{u}_y u_y \Big\rangle,\quad 
    \chi_{dd} = \frac{1}{V} \sum_{x,y} \Big\langle \bar{d}_x d_x \bar{d}_y d_y \Big\rangle,\quad
    \chi_{ud} = \frac{1}{V} \sum_{x,y} \Big\langle \bar{u}_x u_x \bar{d}_y d_y \Big\rangle.
\end{align}
The symmetries of our model imply that $N_u=N_d$ and $\chi_{uu}=\chi_{dd}$. In the above expressions 
the expectation values are defined as
\begin{align}
     \left\langle \mathscr{O} \right\rangle &= \frac{1}{Z} \int [\mathcal{D}\bar u \mathcal{D} u 
     \mathcal{D}\bar d \mathcal{D} d]\hspace{0.1cm} \mathscr{O} \hspace{0.1cm} e^{-S[\bar u,u, \bar d,d]}
\end{align}
with $Z$ being the partition function. The susceptibilities $\chi_{uu}$ and $\chi_{ud}$ diverge as $V$ 
in the FM-phase with non-zero fermion bilinear condensate but saturate in both the PMW phase and the PMS phase.

\section{Fermion Bag Approach}
The traditional approach to solve four-fermion models is by introducing an auxiliary field, where 
we can use the Hubbard-Stratonovich transformation to convert the four-fermion coupling into a fermion 
bilinear by introducing an auxiliary bosonic field \cite{ALF:2020tyi}. Methods of gradient flow, generalizing 
the Lefschetz thimble method to the lattice Thirring model at finite density, have also been recently explored
\cite{Alexandru:2016ejd,PhysRevLett.121.191602}. In this work, we use an alternate approach called the 
Fermion Bag approach \cite{Chandrasekharan:2009wc}. The basic idea of this alternate method is to regroup 
fermion worldlines inside regions denoted as fermion bags. The partition function can then be expressed 
as a sum over configurations of fermion bags. However, since there are several ways to regroup fermion 
worldlines, the approach is not unique and depending on the physics the regrouping must be done thoughtfully. 
When this is possible, this approach can be more efficient than the auxiliary field approach. 

\begin{figure}[H]
	\centering	
	\includegraphics[width=0.75\textwidth]{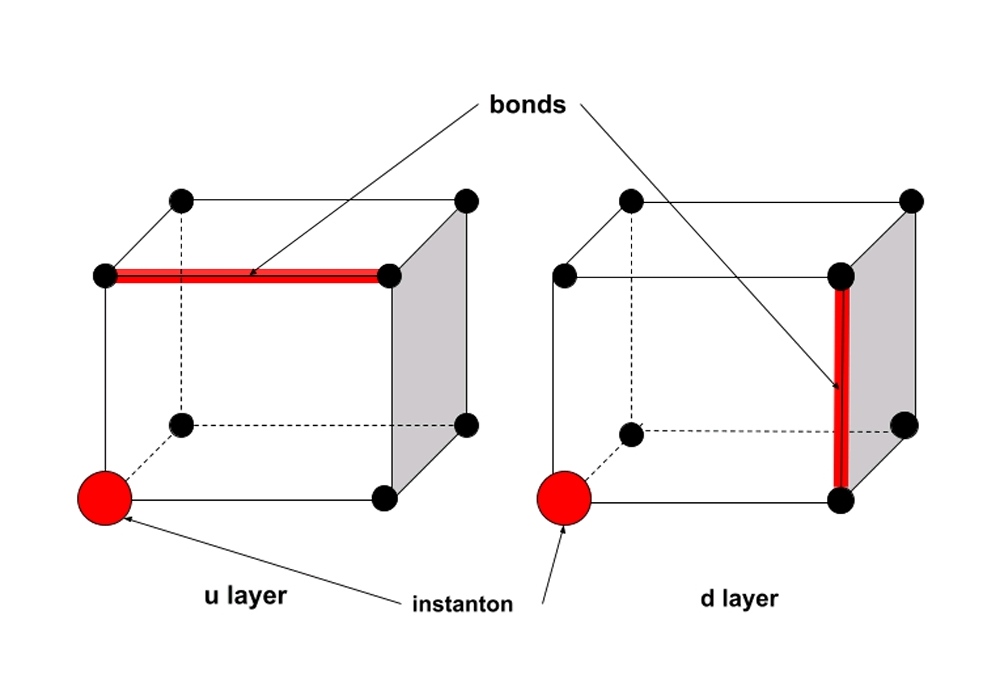}
	\caption{\label{fig:conf} An illustration of a possible configuration $[b,i]$ on a $2^3$ lattice. 
	The red circles denote the instanton sites and red links denote dimers. Dimers on $u$ and $d$ layers 
	can occur on different sites, while instanton positions on both the layers are same and can be 
	visualized as bonds connecting the two layers. The black circles denote free sites and do not contain 
	any instantons or dimers.}
\end{figure}

For our model, several possible ways to identify fermion bags. The first step involves expanding the partition 
function 
\begin{align}
Z = \int [\mathcal{D}\bar u \mathcal{D} u \mathcal{D}\bar d \mathcal{D} d]
   \ e^{-\sum\limits_{x,y}(\bar u_x M_{x,y} u_y +  \bar d_x M_{x,y} d_y  )  + U \sum\limits_{x,y} (\bar u_x u_x \bar u_y u_y +  \bar d_x d_x \bar d_y d_y )+ U^{\prime} \sum\limits_{x} \bar  u_x u_x \bar d_x d_x  } 
\end{align}
as a sum over powers of interactions, as is done in bare perturbation theory. This is done by expanding
\begin{align}
e^{U\bar u_x u_x \bar u_y u_y} = (1 + U \bar u_x u_x \bar u_y u_y) = 
\sum\limits_{b_{x,y} = 0,1}(U\bar u_x u_x \bar u_y u_y)^{b_{x,y}}  \\
e^{U^{\prime} \bar u_x u_x \bar d_x d_x} = (1 + U^{\prime} \bar u_x u_x \bar d_x d_x) 
= \sum\limits_{i_{x} = 0,1}(U^{\prime} \bar u_x u_x \bar d_x d_x)^{i_{x}}.
\end{align}
 We can introduce a new set of variables representing the interactions, which in our case would be 
 \emph{dimers} or \emph{bonds} $b_{x,y}$ for nearest neighbor interactions and \emph{instantons} 
 $i_x$ for single site interactions. Thus, every configuration $[b,i]$ divides all lattice sites into 
 either dimers $[b]$ and $[i]$. Sites that do not belong to a dimer or an instanton will be referred 
 to as free sites. The latter does not  have any instantons or bonds. Figure \ref{fig:conf} illustrates 
 a configuration on a $2^3$ lattice.
 
 One way to identify fermion bags in our model would be to group Grassmann variables that appear at 
 every dimer or instanton and integrate them out. This gives us factors of $U$ or $U^\prime$ for every 
 bond and instanton. The remaining free sites then form their own group of sites and are integrated out 
 separately. The partition function can then be written as a sum over all configurations of $[b,i]$ as 
 \cite{PhysRevD.97.054501},
\begin{align}
Z &=  \sum_{[b,i]} {U^{\prime}}^{N_i} U^{N_u+N_d} {\rm det} (W_u) ~~{\rm det} (W_d)
\end{align}
 where $N_i$ represents the number of instantons, while $N_u$ and $N_d$ represent the number of $u$-dimers 
 and $d$-dimers in the configuration. The matrices $W_u$ and $W_d$ are free staggered fermion matrix 
 restricted to the free sites. We call this approach \textit{strong coupling fermion bag} since in this 
 approach the sizes of the matrices $W_u$ and $W_d$ will be small.

\begin{figure}[tbh]
\begin{center}
\includegraphics[width=0.45\linewidth]{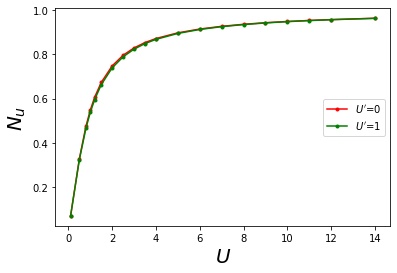}
\includegraphics[width=0.45\linewidth]{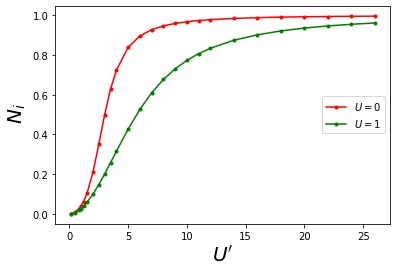}
\caption{\label{fig:ad}The variation of the average bond density ($N_u$) and the average instanton density 
($N_i$) with couplings $U$ and $U^\prime$ for a lattice of linear dimension 2.}
\end{center}
\end{figure}

 An alternate way of constructing the fermion bag is to consider each configuration $[b,i]$ as a term 
 in the perturbative expansion and write the partition function as
\begin{align}
Z = \sum_{[b,i]} U^{'N_i} U^{N_u+N_d} &
\Bigg\{\int [\mathcal{D}\bar u \mathcal{D} u]
e^{-\sum\limits_{x,y}(\bar u_x M_{x,y} u_y)}
\bar{u}_{z_1}u_{z_1}...\bar{u}_{z_k}u_{z_k}
\Bigg\}\nonumber \\
& \times \Bigg\{
\int [\mathcal{D}\bar d \mathcal{D} d]
e^{-\sum\limits_{x,y}(\bar d_x M_{x,y} d_y)}
\bar{d}_{w_1}d_{w_1}...\bar{d}_{w_\ell}d_{w_\ell}
\Bigg\}.
\label{eq:pertexp}
\end{align}
In this expression $k=2N_u+N_i$ and $\ell = 2N_d+N_i$. The terms in the brackets in Eq.~(\ref{eq:pertexp}) can be 
computed using Wick's theorem and summed over all contractions. This yields a simple result
\begin{align}
Z &=  [\mathrm{det}(M)]^2\ \sum_{[b,i]} {U^{\prime}}^{N_i} U^{N_u+N_d} {\rm det} (G_u) ~~{\rm det} (G_d)
\end{align}
where $G_u$ and $G_d$ are $k\times k$ and $\ell \times \ell$ propagator matrix between the sites $z_1,...z_k$ 
and $w_1,...,w_\ell$ respectively. We call this approach the \textit{weak coupling fermion bag}, since the sizes 
of the matrices $G_u$ and $G_d$ will be small at weak couplings. For the efficiency of the fermion bag algorithm,
it is often useful to switch between the two fermion bag formulations.

\begin{figure}[t]
\begin{center}
\includegraphics[width=0.4\linewidth]{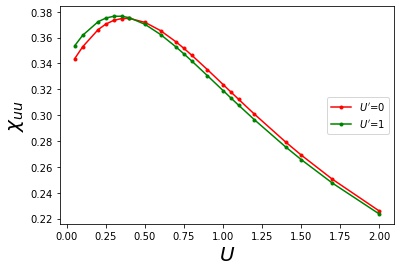}
\includegraphics[width=0.4\linewidth]{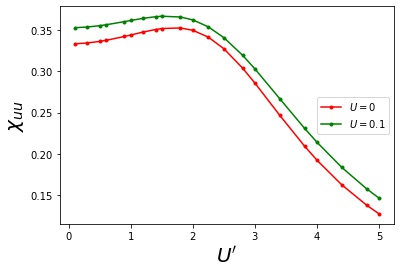}
\includegraphics[width=0.4\linewidth]{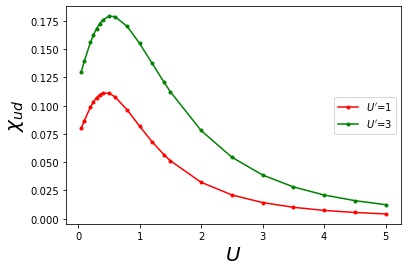}
\includegraphics[width=0.4\linewidth]{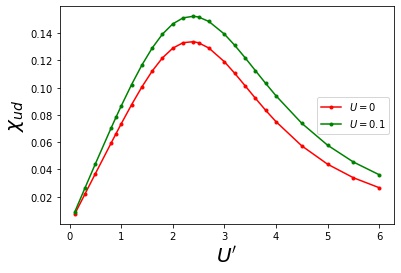}
\caption{\label{fig:spt}The behavior of $\chi_{uu}$ and $\chi_{ud}$ as a function of the couplings $U$ and $U^\prime$.}
\end{center}
\end{figure}

\section{Results}
We are currently developing the fermion bag Monte Carlo algorithm to explore the phase diagram of our model. 
In order to confirm the correctness of the algorithm, it would be useful to compare the results with exact 
calculations. Here we show the results from these exact calculations on a $2^3$ lattice size with 
anti-periodic boundary conditions in all three directions. The behavior of the observables $N_u$ and $N_i$ 
as a function of the couplings $U$ and $U^\prime$ are shown in Fig.~\ref{fig:ad}. These four-point condensates 
are smooth functions, increasing from 0 for small couplings, and approaching 1 for large couplings.

To explore the phase diagram of our model further, it is more natural to study the susceptibilities $\chi_{uu}$ 
and $\chi_{dd}$. These are plotted as functions of the couplings $U$ and $U^\prime$ in Fig.~\ref{fig:spt}. 
In the context of phase transitions, a peak in the plot of the susceptibilities as a function of the coupling 
for a fixed system size, can indicate a  phase transition in the system. However, we must also show that the 
peak actually diverges in the thermodynamic limit to confirm the existence of the phase transition. As
can be seen in Fig.~\ref{fig:spt}, both  susceptibilities show peaks. It is important to point out the fact 
that for very large couplings both susceptibilities as we have defined it becomes small and is related to the 
fact that the lattice is being filled with bonds and instantons. This however does not imply that the 
condensate vanishes. To measure the true condensate, we will need to study the dependence on the system size. 
We expect this to be dramatically different, at large U and small $U^\prime$ as compared to large $U^\prime$
and small $U$. The former will show that $\chi_{uu}$ will grow as a function of the system size while the 
latter will saturate. These can be inferred from previous studies.

\section{Conclusions and Future work}
We have introduced a three-dimensional Gross-Neveu model with two couplings $U$ and $U^\prime$ such that two 
mechanisms of fermion mass generation are possible and can compete. We have shown some exact results for $2^3$ 
lattice and already see some rudimentary features of the phase diagram. In order to understand the physics 
of our model in more detail, we plan to extend our results for larger lattice size using the fermion bag 
algorithm. We hope to explore the phase diagram of our model, and study the nature of the phase transitions. 
If some of these are second order, we plan to calculate the critical exponents.

\bibliographystyle{unsrt}
\bibliography{bibliography.bib}

\end{document}